\documentclass[conference]{IEEEtran}
\IEEEoverridecommandlockouts
\newcommand{\method}[0]{StraTyper}
\newcommand{\naive}[1]{Naive$_{LLM}^{#1}$}
\newcommand{\naivebold}[1]{\textbf{Naive}$_{LLM}^{#1}$}
\newcommand{\methodllm}[1]{\method$_{LLM}^{#1}$}
\newcommand{\methodllmbold}[1]{\textbf{\method}$_{LLM}^{#1}$}
\usepackage{todonotes}
\usepackage[numbers]{natbib}
\usepackage{multirow}
\usepackage{tcolorbox}
\usepackage{subcaption}
\usepackage{graphicx}
\usepackage{booktabs}
\usepackage{tabularx}
\usepackage[linesnumbered]{algorithm2e}
\usepackage{amsmath}
\usepackage{amssymb}
\usepackage{xcolor}
\usepackage{enumitem}
\usepackage{xspace}
\usepackage{hyperref} 
\usepackage{cleveref}
 \usepackage{url}
\usepackage{xcolor}

\newtheorem{definition}{Definition}[section]

\definecolor{commentgreen}{rgb}{0,0.5,0}

\SetCommentSty{mycommfont}

\newcounter{JulianaNOC}
\stepcounter{JulianaNOC}

\newcounter{ChristosNOC}
\stepcounter{ChristosNOC}

\def\BibTeX{{\rm B\kern-.05em{\sc i\kern-.025em b}\kern-.08em
    T\kern-.1667em\lower.7ex\hbox{E}\kern-.125emX}}

\newcommand{\para}[1]{\noindent\textbf{#1.}}
\newcommand{\type}[1]{\texttt{#1}\xspace}
\newcommand{\hide}[1]{}
\begin{document}

\title{\method: Automated Semantic Type Discovery and \\ Multi-Type Annotation for Dataset Collections}

\author{\IEEEauthorblockN{Christos Koutras}
\IEEEauthorblockA{
\textit{New York University}\\
New York, NY, USA \\
christos.koutras@nyu.edu}
\and
\IEEEauthorblockN{Juliana Freire}
\IEEEauthorblockA{
\textit{New York University}\\
New York, NY, USA \\
juliana.freire@nyu.edu}
}

\maketitle

\begin{abstract}
Understanding dataset semantics is crucial for effective
search, discovery, and integration pipelines. For this purpose, column-type annotation (CTA) methods associate columns in tabular datasets with semantic types that accurately describe their contents, using pretrained deep learning models or Large Language Models (LLMs). However, existing approaches have important limitations. First, they require users to specify a closed set of semantic types either during training or inference, which hinders their application to domain-specific datasets where predefined type vocabularies often lack adequate coverage and specificity. They also assume that each column is associated with a single semantic type, an assumption that does not hold for many real-world datasets. Lastly, while proprietary LLMs have shown effectiveness for CTA, they incur high monetary costs and produce inconsistent outputs across similar columns, leading to type redundancy that degrades downstream applications.

To address these challenges, we introduce \method{}, a cost-effective method for \emph{column type discovery} (CTD) and \emph{multi-type annotation} (CMTA) in dataset collections. StraTyper eliminates the need for predefined semantic labels by systematically employing LLMs to discover types tailored to the target dataset collection. Through strategic column clustering, controlled type generation, and iterative cascading discovery, StraTyper balances type precision with annotation coverage while minimizing LLM costs. Our experimental evaluation, both manual and LLM-assisted, on real-world dataset collections demonstrates that StraTyper discovers accurate types for both numeric and textual data, achieves substantial cost savings compared to commercial LLMs, and effectively handles multi-typed columns. We further show that StraTyper's annotations improve downstream tasks, including join discovery and schema matching, outperforming LLM-only baselines.
\end{abstract}


\section{Introduction}
\label{sec:intro}

Semantic type annotation is fundamental to making tabular data discoverable, understandable, and usable across modern data ecosystems. Unlike syntactic types (e.g., integer, string) that merely describe data representation, semantic types capture the real-world meaning of column contents, e.g., distinguishing whether a string column contains a New York City school code (district borough number, i.e., DBN), the name of a borough, or a city. This semantic understanding is critical for downstream data management tasks, including dataset discovery in large repositories, data integration, data cleaning, and knowledge graph enrichment~\cite{freire2025large,rahm2001survey,knowledge-graph-construction-acmsur2024}. 

Real-world datasets, particularly domain-specific collections, require semantic types that are both specific and comprehensive. Generic ontologies such as DBpedia may classify a column as ``Educational Institution", but downstream applications often require the precision of ``NYC Public High School" to avoid retrieving irrelevant data about colleges, libraries, or medical centers. This gap becomes acute in specialized domains and
repositories, such as NYC OpenData~\cite{nycopendata}, which contain thousands of datasets with localized types (e.g., NYC agencies, boroughs, public schools). Manual annotation of such fine-grained types is impractical at scale.

\para{Column Type Annotation Approaches: Limitations} CTA methods attempt to address this challenge by automatically annotating columns with semantic types. However, learning-based CTA methods, whether deep learning models trained from scratch or fine-tuned transformer-based language models like BERT, face fundamental scalability and applicability constraints~\cite{feuer2024archetype}. These approaches \emph{require large volumes of annotated training data} to achieve acceptable performance, creating prohibitive data-labeling costs
and limiting their applicability to rare types.
%
They are also \emph{constrained to a closed set of semantic types fixed at training time} and cannot recognize new types encountered in the wild. This limits their utility in real-world settings where datasets are diverse and seldom map cleanly onto pretrained categories. Even when column types match their training labels, learning-based CTA models degrade substantially under \emph{distribution shift}, i.e., when evaluation data differs from training data~\cite{feuer2024archetype}, and therefore generalize poorly beyond their training corpus.

\begin{figure}[t!]
    \centering
    \includegraphics[width=\columnwidth]{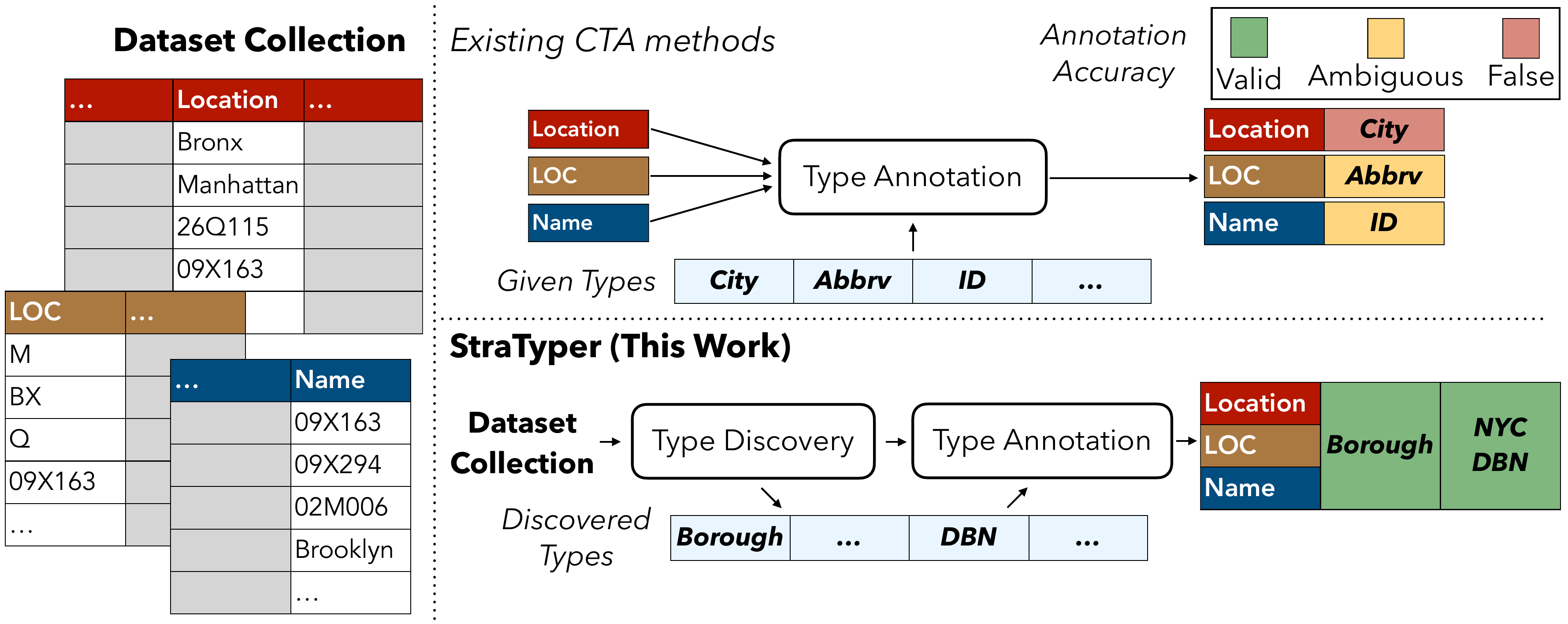}
    \caption{\method{} vs. CTA methods: CTA methods assign each column one type from a predefined list. For columns Location, LOC, and Name, they assign \type{City}, \type{Abbrv}, and \type{ID} which are incorrect, generic, or ambiguous types that fail to capture multi-type values. In contrast, \method{} \emph{discovers} collection-specific types and consistently annotates similar columns, correctly assigning \type{Borough} and \type{DBN} to all three columns.}
    \label{fig:motivation}
    \vspace{-.45cm}
\end{figure}

Recent advances in large language models (LLMs) have opened new possibilities for CTA. 
Trained on vast, diverse text corpora, LLMs accumulate knowledge spanning a broad range of semantic types, and can perform zero-shot CTA by classifying columns against a user-provided set of candidate types at inference time, without task-specific training data~\cite{feuer2024archetype,kayali2024chorus,korini2023column}.
%
For instance, given the value ``Stuyvesant", GPT-3.5-Turbo 
correctly identifies it as a ``High School in New York City'', showing both semantic understanding and adaptability to domain-specific types.
However, these methods still require users to provide a predefined set of candidate semantic types, and this  poses multiple challenges: 
\emph{coverage} (Challenge 1), as users may be unaware of all semantic types present in a  collection, particularly in large, heterogeneous data repositories or unfamiliar domains;
\emph{specificity} (Challenge 2), since users must choose a  granularity for each type (e.g., ``Educational Institution'' vs. ``NYC Public High School'') without knowing the data; and 
\emph{scalability} (Challenge 3), as manually curating comprehensive type sets is impractical for large collections with diverse schemas.
In addition, because these methods assign a single type to each column, they \emph{cannot capture columns whose values straddle multiple types} (Challenge 4). \Cref{fig:motivation} illustrates these limitations: CTA methods cannot correctly map predefined types to columns and fail to capture the presence of multiple types.

\para{Our Approach: Semantic Type Discovery}
To overcome these challenges, we introduce \method{}, a cost-effective framework for \emph{Column Type Discovery (CTD)} and \emph{Column Multi-Type Annotation (CMTA)}.
%
Unlike existing approaches, which require the type vocabulary to be specified at training 
or inference time, 
\method{} uses LLMs to automatically discover semantic types tailored to the collection at hand.
By prompting an LLM over clusters of similar columns and consolidating the types it discovers, \method{} generates domain-specific vocabularies that capture the coverage and specificity required for accurate, reliable annotation (Challenges 1, 2).
%
Users no longer need prior knowledge of the collection's semantics or manually curated type sets (Challenge 3), and processing clusters rather than individual columns substantially reduces the cost of using proprietary LLMs.
%
\method{} also handles multi-typed columns (Challenge 4), whose values belong to multiple semantic types.
As \Cref{fig:motivation} shows,  \method{} discovers the types \type{Borough} and \type{DBN} that correspond to New York City boroughs and public school identifiers, respectively, and it also recognizes that the columns Location, LOC, and Name contain values of both types.

Using LLMs for CTD introduces a new challenge: while LLMs can infer the semantic type of a given column, they may assign different labels to semantically equivalent columns, leading to \emph{type redundancy} (Challenge 5). This can negatively affect downstream applications, such as semantic join discovery~\cite{dong2023deepjoin, koutras2025omnimatch, guo2025snoopy}, that rely on consistent
annotations. \method{} controls type generation and avoids redundancy by maintaining an inverted index of discovered types, which it reuses across columns.

\para{Summary of Contributions} 
\method{} is, to our knowledge, the first framework designed for automated semantic type discovery in dataset collections. Its design follows a consistent pattern: each stage trades away coverage or generality for precision where errors would otherwise propagate uncorrected, and a paired low-cost mechanism recovers what was sacrificed. This pattern, rather than any single technique, enables StraTyper to achieve effectiveness, cost-efficiency, and robustness simultaneously, as we validate in Section~\ref{sec:experiments}. Our main contributions can be summarized as follows:
%

\begin{itemize}[leftmargin=*]
\item 
\emph{We design a clustering and indexing strategy that derives pure clusters and
makes open-vocabulary discovery tractable at the collection level.} Clustering columns by name and value yields pure clusters that can be batch-processed, while a value-type inverted index reuses discovered types across clusters. Together, they substantially reduce LLM usage, enable multi-type inference, and prevent the type redundancy that independent per-column discovery would produce (\Cref{sec:semantic_discovery}). 
\item \emph{We develop a Chain-of-Thought prompting strategy} that decomposes closed-set annotation into sequential verification steps across targeted candidate-type lists, achieving reliable performance with open-source LLMs at low cost (\Cref{sec:annotation}).
\item \emph{We propose an iterative cascading mechanism} that recovers the coverage sacrificed by  purity-first clustering: it discovers types at strict similarity thresholds and dynamically propagates them as candidates to broader column groups, effectively balancing precision and coverage, removing the need for manual threshold tuning (\Cref{sec:iterative-cascading}).
\item \emph{We demonstrate through extensive experiments} that \method{} matches the annotation quality of state-of-the-art proprietary models at a substantially lower 
cost, while delivering higher utility for downstream tasks such as join discovery and schema matching (\Cref{sec:experiments}).
\end{itemize}
\section{Problem Setting \& Our Approach}

\para{Problem Setting and Definition} 
We address the problem of discovering semantic types and annotating table columns with one or more types that accurately describe their instance values. 
%
CTA approaches \cite{hulsebos2019sherlock, zhang2020sato, deng2021turl, miao2023watchog, suhara2022annotating,korini2023column, kayali2024chorus,li2024table, feuer2024archetype, wei2024racoon, ding2025retrieve, xiao2025cents} focus on  annotation, and select from a set of predefined labels the type for a given column:

\begin{figure*}[t]
    \centering
    \includegraphics[width=.75\textwidth]{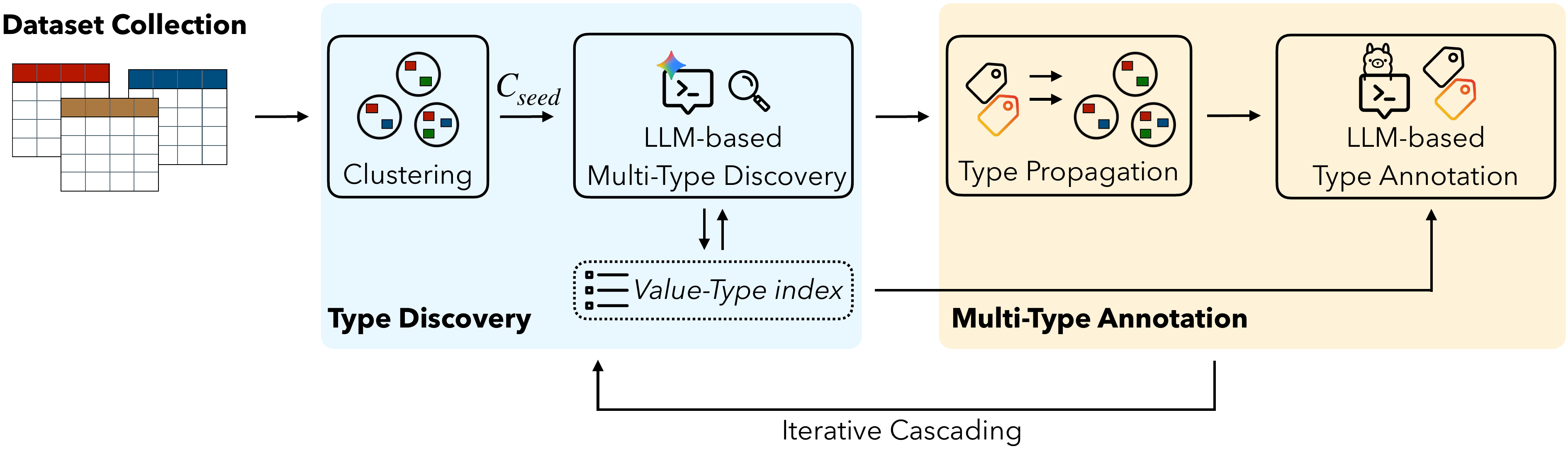}
    \caption{\method{} receives a dataset collection as input and proceeds to open-type discovery and multi-type annotation, either in a single-pass or an iterative cascading manner.}
    \label{fig:overview}
\end{figure*}

\begin{definition}[Column Type Annotation (CTA)]
Given a table $T$ and a column $c_i$ in $T$, a CTA model M with type vocabulary $M_{type}$ assigns a single column type $t \in M_{type}$ that best
describes the semantics of $c_i$.
\end{definition}


\noindent We consider a different problem that has two complementary goals. 
Given a dataset collection $\mathcal{D} = \{T_i, i \in [1, n]\}$, where each table $T_i$ comes with a set of columns $\mathcal{C}_i$,
the first goal is to discover semantic types that capture concepts aligned with a given dataset collection:

\begin{definition}[Column Type Discovery (CTD)]
CTD identifies a set of 
types $\mathcal{S}$ that accurately describe columns across datasets, i.e., $\forall T_i \in \mathcal{D}, c \in \mathcal{C}_i, \exists S_i \subseteq \mathcal{S}: Types(c) \equiv S_i$.
\end{definition}

\noindent Using the discovered types, the second goal is to assign one or more types to each column to properly capture the presence of values from different domains: 

\begin{definition}[Column Multi-Type Annotation (CMTA)]
CMTA associates to each column $c\in\mathcal{C}_i$  one or more semantic types from $\mathcal{S}$, i.e., $\forall T_i \in \mathcal{D}, c \in \mathcal{C}_i: CMTA(c) = \{s_1, \dots, s_d\} \subseteq \mathcal{S}$.
\end{definition}


\para{\method{} at a Glance} \method{} solves CTD  and CMTA through the steps summarized in \Cref{fig:overview}, 
combining semantic clustering, controlled type generation, and the selective use of open-source models to achieve a cost-effective solution.


\noindent-- \emph{Column Clustering (Section \ref{sec:clustering})}: \method{} clusters columns separately by name and by value, and forms \emph{seed clusters} from those grouped together under both. A single type-discovery call then serves all members of a seed cluster, which reduces LLM usage and prevents semantically identical columns from yielding redundant, synonymous types. 

\noindent-- \emph{LLM-based Multi-Type Discovery (Section \ref{sec:discovery})}: Given seed clusters, \method{} leverages a commercial LLM to discover semantic types tailored to the collection. To avoid redundancy, it reuses previously discovered types through a value-type inverted index (\emph{values}$\rightarrow$\emph{semantic types}), and it samples values to surface the multiple types of a column. The sampling strategy is modular; stratifying values by string length works well in our experiments (Section~\ref{sec:ablation}), while other strategies may work better in other settings.
%

\noindent-- \emph{Multi-Type Annotation (Section \ref{sec:annotation})}: \method{} propagates the discovered types to the columns of each seed cluster and annotates the remaining columns with open-source (OS) LLMs, avoiding the cost of proprietary models. Since OS models are less capable, \method{} constrains their search space to a small set of high-confidence candidates, retrieved via clustering and the index, guiding them with a step-by-step verification prompt.


\noindent-- \emph{Iterative Cascading Discovery and Annotation (Section \ref{sec:iterative-cascading})}:  A fixed similarity threshold either misses similar columns (if too high) or creates noisy clusters (if too low). \method{} therefore iterates over decreasing thresholds, discovering precise types at high thresholds and propagating them to other
columns at lower ones. Discovery is invoked only when propagation fails, which balances type precision against LLM cost and annotation coverage.

Each design choice is empirically validated in Section~\ref{sec:experiments}: structured prompting and type reuse (Section~\ref{sec:evaluation}), open-vocabulary discovery (Section~\ref{ssec: expsetup}), and seed cluster precision, length-stratified sampling, iterative cascading, and model choices (Section~\ref{sec:ablation}).



\section{Discovering Semantic Types}
\label{sec:semantic_discovery}
Seed clustering must produce \emph{pure} clusters,  
{i.e., every pair of columns in a cluster shares at least one semantic type, because a single discovery prompt is applied uniformly to the cluster and an impure cluster can corrupt the annotation of every column in it. Purity matters more than recall here, since columns missed by clustering are recovered later (Sections~\ref{sec:annotation} and~\ref{sec:iterative-cascading}).
Name-based signals alone are insufficient, since syntactically similar headers can denote unrelated concepts (e.g., two ID columns from different domains); value-based signals are also insufficient, since columns with similar distributions can be semantically unrelated (e.g., a percentage and a normalized score both ranging from 0 to 100). 
We therefore intersect metadata (column names) and content (column values) signals, which empirically yields pure clusters (\Cref{tab:seed_precision}) without a supervised similarity model. Grouping columns this way also lets us discover types for a whole cluster with a single prompt, reducing token usage and preventing the synonymous types that independent per-column processing would produce.


\para{Metadata Clusters ($\mathcal{C}_{name}$)} We generate embeddings for all column headers using Sentence-BERT (SBERT)~\cite{reimers2019sentence} and group them using the \emph{Fast Community Detection} algorithm from the sentence-transformers package~\cite{sbert}. We opt for this algorithm as it determines clusters via a similarity threshold ($\tau_{name}$), that defines the minimum cosine similarity for two column names to be grouped together, rather than requiring a predefined cluster count.
This naturally clusters synonymous headers (e.g., \texttt{Location} and \texttt{LOC}).

\para{Content Clusters ($\mathcal{C}_{value}$)} To capture semantic similarity based on column values, we apply the same clustering logic with a separate threshold ($\tau_{value}$). However, we differentiate column embeddings for textual and numerical data as follows:

\noindent-- \emph{Textual Columns}: We compute the average SBERT embedding of the unique values stored in each column.  
Unlike traditional overlap metrics (e.g., Jaccard similarity), this embedding-based approach captures latent semantic relationships, grouping columns with the same type(s) even when their value sets are disjoint.

\noindent-- \emph{Numerical Columns}: Since language models cannot effectively capture numerical distributions, we adopt a statistical clustering approach for numerical columns. Each column is represented by the concatenation of two feature groups: (i) a discretized histogram density vector with $b$ bins computed over Z-score normalized values to capture distributional shape, and (ii) a set of scale-sensitive magnitude features computed on the raw values. 
The latter consist of five signed log-percentiles ($\text{sign}(p) \cdot \log(1+|p|)$ for $p \in \{p_5, p_{25}, p_{50}, p_{75}, p_{95}\}$), a binary indicator for the presence of negative values, and the fraction of integer-valued entries. Intuitively, the histogram captures distributional shape invariant to scale, while the magnitude features preserve the absolute scale and granularity signals that Z-score normalization would otherwise discard, preventing columns with similar shapes but vastly different ranges (e.g., percentages vs.\ population counts) from being clustered together.


\para{Discovery Seed Clusters ($\mathcal{C}_{seed}$)} Finally, we construct seed clusters for type discovery by intersecting the metadata-based and content-based clusters. 
Two columns are in the same seed cluster when they are co-located in the same name-based and value-based clusters. This strict intersection requirement acts as a high-precision filter, ensuring that grouped columns are semantically coherent. While we employ a unified threshold ($\tau_{name} = \tau_{value} = \tau$) in our baseline evaluation, this framework allows for domain-specific tuning; for instance, lowering $\tau_{name}$ in datasets with noisy headers to rely more heavily on value-based signals.

\subsection{LLM-based Type Discovery}
\label{sec:discovery}

\method{} synthesizes a tailored prompt for each seed cluster to facilitate zero-shot type discovery. Since naive prompts may not handle open-world, multi-type scenarios effectively, our prompt construction pipeline incorporates three components: representative value sampling, context injection, and dynamic type retrieval.

\para{Value Sampling for Multi-Type Discovery}
Discovering multiple types in a cluster requires that the LLM see examples of each. Random or frequency-based sampling can miss this, since rare types are easily overshadowed by dominant ones. \method{} treats value sampling as a modular step; here we use a lightweight, structure-based strategy, \emph{length-stratified sampling}, and show that it makes multi-type discovery feasible at negligible cost. Other strategies, including semantic ones, can be plugged in. The strategy builds on the observation that distinct semantic types often exhibit characteristic string lengths (e.g., NYC DBNs have a fixed length, whereas borough names vary). Given a sampling budget $k$, we construct a representative value set $V_{rep}$ as follows. First, we aggregate all unique values from seed cluster $\mathcal{C}_{seed}$ and partition them by string length. Second, we sort these groups by size in descending order. When the number of groups $M$ exceeds $k$, we select one representative from each of the $k$ most populated groups, favoring structural breadth. When $k > M$, we sample from all groups and allocate additional samples to the most populous ones to capture diversity within identical structures. The strategy cannot separate types whose values have similar lengths (e.g., person names mixed with place names); in that case it reduces to sampling within a single group. Section~\ref{sec:ablation} evaluates it against clustering-based alternatives on the multi-typed columns of NYC-E, where length aligns well with type boundaries.

\para{Context Injection} Aggregating columns from different tables into a single cluster strips their original table context, which is valuable for CTA tasks~\cite{zhang2020sato}. This particularly harms numerical columns with opaque names, where neighboring columns provide critical disambiguation. For instance, ``Count'' could represent population, transactions, or inventory depending on table context. To restore this context, we compute a \emph{header co-occurrence profile} for each cluster by identifying the top-$N$ column names most frequently appearing in the same tables as cluster columns. These are injected into the prompt, providing the LLM necessary relational context for semantic disambiguation.

\begin{figure*}[t!]
\centering
\footnotesize
\begin{tabularx}{\textwidth}{X@{}}
\toprule \bottomrule
\textbf{Type Discovery Prompt (Example)} \\
\toprule \bottomrule
You are given a set of columns grouped together based on embedding similarity of the values they store and their column names. Their column names, a representative sample of values, and a set of possible semantic types are provided. You are also given information about columns that co-appear with the ones in the cluster. Your task is to identify all distinct semantic types that accurately describe the data within a group of columns. \\ 
\addlinespace

\textbf{Column names in cluster:} \texttt{[Location, LOC]} \\ 
\addlinespace

\textbf{Sample Values:} \texttt{Manhattan | Bronx | M | Q | 09X163 | 02M006} \\ 
\addlinespace

\textbf{Context Columns:} \texttt{Borough Number (2/2) | Building Address (1/2)} \\ 
\addlinespace

\textbf{Possible Semantic Types:} \texttt{["NYC Borough", "NYC Neighborhood", ...]}  \\ 
\addlinespace

Your analysis should follow these key rules: 

\begin{itemize}
\item[--] \emph{\textbf{Be specific:}} Semantic types must be more precise than generic data types (e.g., string, integer). They should capture the intended use of the data.
\item[--] \emph{\textbf{Use all available information:}} Analyze both the column names and the sample values. Expand any abbreviations to inform your choice. If names and values conflict, prioritize what the values represent.
\item[--] \emph{\textbf{Reuse or Propose:}} For EACH distinct concept you identify in the cluster, you must follow this logic:
\begin{enumerate}
    \item[1.] \emph{\textbf{Check for Reuse:}} Look at the ``Possible Semantic Types'' list. If an existing type exactly and unambiguously describes that concept, you MUST use it.
    \item[2.] \emph{\textbf{Propose New:}} If no existing type from the list is an exact match for that specific concept, you MUST propose a new, specific semantic type for it.
\end{enumerate} 
\item[--] \emph{\textbf{Do not select a ``closest match'':}} This is critical. If an existing type is only similar but not an exact fit for a concept, you must propose a new type instead.
\item[--] \emph{\textbf{Consider the broader context:}} Columns that co-appear with the ones in the cluster can be valuable context to better understand semantic types. Use the context provided to inform your choice.
\item[--] \emph{\textbf{Avoid redundancy}}: Your final list of types should be distinct and not overlap.
 
\end{itemize}
Output ONLY a JSON dict \texttt{\{"answer": [semantic types]\}}. \\ 

\toprule \bottomrule
\end{tabularx}

\caption{Type Discovery Prompt, integrating column names in seed cluster, length-stratified value samples, context based on co-occurrence profiles, dynamically retrieved possible semantic types and detailed instructions, as discussed in \Cref{sec:discovery}.}
\vspace{-0.3cm}
\label{fig:prompt_structure}
\end{figure*}
\subsection{Column Clustering}
\label{sec:clustering}

\para{Dynamic Type Retrieval} A significant challenge in open-type semantic discovery is the generation of redundant, synonymous types (e.g., ``NYC Borough Name'' and ``Borough'') that represent identical concepts across clusters. Naively providing the LLM with the complete set of previously discovered types is prohibitively expensive in terms of token usage and also increases the likelihood of hallucinations or inaccurate annotations \cite{feuer2024archetype}. 
To address this challenge, we implement a dynamic type retrieval mechanism using an inverted index that maps each unique value to its associated semantic types. The mechanism operates as follows:

\noindent-- \emph{Index Update:} As semantic types are discovered for a cluster, we propagate these labels to all unique values within that cluster and update the global inverted index accordingly.

\noindent-- \emph{Retrieval:} When processing a new seed cluster, we query the index using the unique values stored across its columns. If any values have associated types from prior iterations, we retrieve and present those types to the LLM as candidate semantic types, indicating that these are known, validated labels.

\noindent-- \emph{Fallback:} If no values are found in the index, we provide a filtered subset of previously discovered types that match the cluster's data characteristics (i.e., numerical types for numerical columns, textual types for textual columns). To mitigate the risk of large candidate sets, we process seed clusters in descending order of size. This ensures the inverted index is populated quickly with the most prevalent values first, significantly reducing the likelihood of injecting prompts with larger candidate sets. This ordering affects only the granularity of the discovered vocabulary, since the LLM always independently decides whether an available candidate suffices or a new type should be proposed, regardless of processing order.

\para{Candidate Set Initialization} The inverted index is initially empty, so the first cluster is processed without candidate types and the LLM discovers freely from column names, values, and context alone, while subsequent clusters benefit from the incrementally populated candidate set. In a symmetric manner, a user-supplied vocabulary can seed the index so that predefined types enter discovery prompts as initial candidates, in which case the LLM reuses them whenever they adequately describe a cluster and proposes new types only for concepts the vocabulary does not cover. \method{} therefore supports closed-vocabulary annotation as a natural special case, and since LLM-based CTA has been shown to be competitive with trained methods on closed-vocabulary benchmarks \cite{feuer2024archetype}, we expect the cost of this generality to be modest.

\para{Structured Prompt Synthesis} Finally, we integrate the value samples, context, and retrieved types into a structured prompt, as shown in \Cref{fig:prompt_structure}. The prompt guides the LLM through a reasoning path for discovering semantic types for columns in the seed cluster by explicitly instructing it to: 1) consider column names, value samples, and provided context, while prohibiting the generation of generic data types (e.g., \texttt{String}, \texttt{Integer}), 2) minimize redundant type creation by adopting a strict decision logic using the retrieved semantic types, i.e., the LLM must first evaluate whether an existing candidate type provides an exact match, and generate a new type only when no existing candidate type is semantically sufficient to cover any of the observed concepts, 3)  disallow closest-match approximations to ensure precision in discovered types, 4) constrain the output to a strict JSON format containing only the final semantic labels to enable automated parsing.

\section{Type Propagation and Annotation}
\label{sec:annotation}

After the discovery phase, \method{} annotates columns in the dataset collection using the discovered semantic types. A naive approach that annotates each column individually against the full set of discovered types is prone to hallucinations as the candidate list grows large. 
We address this through a two-tier strategy: high-confidence batch-type propagation for seed clusters (\Cref{sec:seed-cluster-propagation}) and targeted, closed-set annotation using open-source LLMs for the remaining columns (\Cref{sec:targeted_annotation}). 

\subsection{Seed Cluster Type Propagation}
\label{sec:seed-cluster-propagation}

Seed clusters ($\mathcal{C}_{seed}$) are highly coherent by construction (Section \ref{sec:clustering}), so we propagate their discovered types to the member columns. 
When a cluster has multiple types, however, a homogeneous column should not receive all of them, whereas a genuinely multi-typed column should. We therefore apply a lightweight heuristic with two signals: string-length distribution and numeric/textual purity. Since values of distinct types often differ in format, and hence in length, we flag a column as multi-typed if there is a gap between consecutive sorted unique string lengths that exceeds the column's interquartile range (IQR), or if its values mix numeric and textual content. Columns that raise no flag are treated as homogeneous and routed to closed-set annotation (Section~\ref{sec:targeted_annotation}), which selects the most fitting type(s) from the cluster's discovered set using an open-source LLM. Flagged columns retain the full set of discovered types. This avoids over-labeling homogeneous columns while keeping inference calls low. Like length-stratified sampling, the heuristic detects only types that differ in length or numeric/textual nature; other signals can be substituted.

\subsection{Targeted Closed-Set Annotation}
\label{sec:targeted_annotation}

Since seed clusters are formed through a strict intersection of name-based and value-based similarity (Section \ref{sec:clustering}), they may fail to capture columns that share the same semantic types but considerably differ either in their column names or in their values. As a result, a potentially large number of columns fall outside the seed clusters and require a dedicated annotation mechanism. To keep this step cost-effective at scale, we perform closed-set annotation using an open-source LLM through carefully scoped candidate sets and structured prompting, as described below.

\para{Candidate Set Retrieval} One possible approach to label the remaining columns is to prompt the LLM using the set of discovered types. However, the effectiveness of LLMs has been shown to decrease with prompt size and number of classification choices~\cite{feuer2024archetype}.  Therefore, instead of overwhelming the LLM with the full set of discovered types, we construct targeted candidate sets for each column. We derive these candidates from three complementary signals created during the clustering phase:

\noindent-- \emph{Name-Based Cluster Candidates}: We examine the column name-based cluster ($\mathcal{C}_{name}$) containing the target column, and identify any columns within that cluster that have already received annotations from the discovery phase. Their types are added to the candidate set, leveraging the assumption that columns with similar headers likely share semantic types.

\noindent-- \emph{Value-Based Cluster Candidates}: Similarly, we examine the value-based cluster ($\mathcal{C}_{value}$) containing the target column and aggregate types from any already-annotated column. This captures columns that have different names but contain semantically similar values.

\noindent-- \emph{Inverted Index Lookups for Candidates}: 
We query the Value-Type index using the target column's unique values. If the values appear in the index, we retrieve the associated types.

\para{Structured Prompting Strategy} After scoping down the candidate sets, we employ an open-source LLM for closed-set annotation. However, simply prompting the LLM to select from candidate types would fail to yield reliable results for multi-type annotation, particularly because open-source models often lack the ability to perform implicit multi-step reasoning. To address these limitations, we design a structured prompt that decomposes the closed-set multi-type annotation task into sequential verification steps. By explicitly serializing the verification process, we induce a Chain-of-Thought~\cite{wei2022chain} that scaffolds the model's reasoning.

The prompt receives as input: (i) the target column's name, (ii) a sample of its table context (column names and top-$l$ rows), and (iii) a length-stratified sample of its values (using the method from \Cref{sec:discovery}). 
The LLM iterates through the candidate set and verifies whether each candidate fits the semantics of the target column. The verification logic differs by column data type:
For \emph{textual data}, the LLM inspects whether each candidate type captures the semantics of any value sample, i.e., a candidate is accepted if it describes at least a subset of the observed values;  for \emph{numerical data}, the decision relies 
on table context and column name, since numerical values alone are often insufficient to determine semantics. 
The prompt explicitly instructs the LLM to reject all candidates if none provides an accurate match. This structured approach enables multi-label annotation while preventing forced assignments and enabling the model to abstain when confidence is low, maintaining annotation precision.

\para{Type Aggregation} 
The annotation strategy varies by type, reflecting fundamental differences in how semantic ambiguity manifests:

\noindent-- \emph{Textual Columns}: Since textual columns may 
contain values from multiple semantic types, we treat the three candidate sources (name-based and value-based clusters, and the inverted index) as complementary views. We perform closed-set annotation for each source independently and union the outputs. 
This 
keeps individual candidate sets small while ensuring comprehensive coverage and enables the discovery of all types present in multi-typed columns.

\noindent-- \emph{Numerical Columns}: Numerical distributions are often ambiguous, since columns sharing identical value ranges might represent entirely different concepts (e.g., percentages and normalized scores both range from 0 to 100). Consequently, we restrict candidate retrieval exclusively to name-based clusters, bypassing value-based clusters and index lookups.  This conservative approach prevents spurious semantic associations that could arise from coincidental numerical overlap, instead relying on the stronger signal of column naming conventions and table context.
\section{Iterative Cascading Discovery}
\label{sec:iterative-cascading}

\begin{figure}[t!]
    \centering
    \includegraphics[width=\columnwidth]{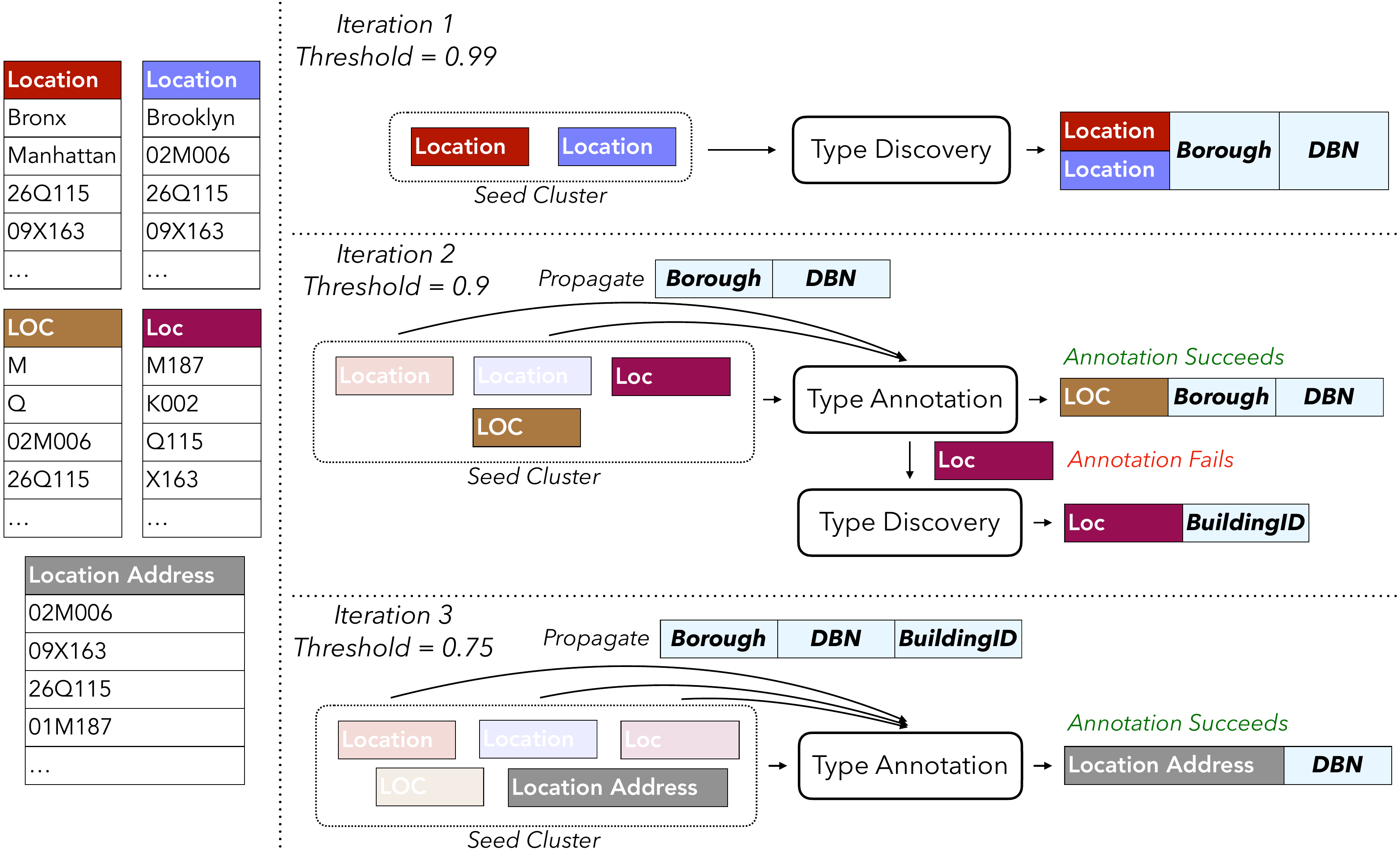}
    \caption{Iterative Cascading discovery: \emph{i}) \method{} discovers semantic types from seed clusters at a high threshold ($\tau=0.99$), \emph{ii}) propagates them to sibling columns in lower threshold clusters (LOC and Location Address for $\tau \in \{0.9, 0.75\}$) via closed-set annotation, \emph{iii}) discovers types for residual unannotated columns (Loc for $\tau=0.75$).} \vspace{-.3cm}
    \label{fig:ic}
\end{figure}

Relying on a single similarity threshold presents a fundamental trade-off: high thresholds yield clean, coherent clusters with low coverage, while low thresholds maximize coverage but risk grouping mixed-type columns, producing overly generic or noisy types. To eliminate manual tuning and balance semantic precision with broad coverage, \method{} employs an \emph{Iterative Cascading} (IC) framework. \Cref{fig:ic} illustrates IC through an example.


\para{The Cascading Process} IC operates over a predefined sequence of decreasing similarity thresholds. Rather than treating these as separate independent runs, we cascade the knowledge (i.e., semantic type annotations) from stricter iterations (higher thresholds) to inform and constrain type discovery in subsequent looser iterations (lower thresholds). At each similarity level $\tau_{sim}$, IC performs the following operations:

\noindent-- \emph{Seed Cluster Formation}: We generate seed clusters using threshold $\tau_{sim}$ (\Cref{sec:clustering}). Initial iterations produce small, dense, semantically pure clusters due to higher similarity thresholds. Subsequent iterations yield progressively larger, looser clusters that capture more subtle semantic relationships.

\noindent-- \emph{Candidate Type Capture}: For each seed cluster, we identify columns previously annotated in higher-threshold iterations. These columns are excluded from further processing, while their discovered types are aggregated to form a high-confidence candidate set for the remaining unannotated columns within the cluster.

\noindent-- \emph{Targeted Closed-Set Annotation}: 
Before invoking type discovery, IC attempts to annotate remaining columns using types from their already-annotated cluster siblings. 
We employ the open-source LLM for closed-set annotation from \Cref{sec:targeted_annotation}, leveraging columns annotated at high thresholds to label columns that qualify only for a seed cluster at lower thresholds. This approach uses zero-cost open-source LLMs, maintains small, relevant candidate sets, and ensures type consistency.

\noindent-- \emph{Discovery on Residual Columns}: Columns that remain unannotated after closed-set annotation are forwarded to LLM-based type discovery (\Cref{sec:discovery}). These columns are semantically distinct from their cluster neighbors and would be falsely annotated if we relied solely on type propagation.
By invoking the discovery LLM only when novel semantic types must be discovered, we avoid erroneous type propagation while limiting token usage.

\para{Residual Column Annotation} Following the cascade process, any remaining unannotated columns undergo a final iterative annotation using the closed-set strategy from Section \ref{sec:targeted_annotation}. To maintain precision, we proceed hierarchically: first attempting annotation using candidate type sets extracted from name-based and value-based clusters at higher thresholds, then progressively considering lower thresholds. This ensures that \method{} evaluates whether fine-grained semantic types fit the remaining columns before resorting to more general types.

\para{Benefits of Iterative Cascading} The cascading approach provides two key advantages. First, it \emph{balances precision and annotation coverage} by capturing fine-grained semantics in early iterations and propagating them to broader groups of columns, thereby avoiding generic type assignments that arise from clustering at lower similarity thresholds alone. For instance, \Cref{fig:ic} shows that relying solely on a loose threshold $\tau=0.75$ would create an impure seed cluster, leading to either discovering overly generic types (e.g., ``Location'') or propagating incorrect types to every column in the cluster. Second, the framework \emph{achieves computational efficiency by minimizing expensive discovery LLM calls}. Columns receive types through propagation and targeted annotation whenever possible, with discovery invoked only for semantically novel cases. This reduces overall token usage and associated costs while maintaining annotation quality.

\section{Experimental Evaluation}
\label{sec:experiments}

We conduct a detailed evaluation of \method{} to assess: \emph{i}) token usage and monetary costs, \emph{ii}) quality of the derived semantic types, \emph{iii}) cost-effectiveness tradeoffs, \emph{iv}) utility
for downstream tasks, including join discovery and schema matching tasks. We also perform ablations across key parameters and components of our method. Our main findings are:

\begin{itemize}[leftmargin=*]
    \item Cost Efficiency: \method{} reduces costs incurred by LLM usage by up to 2x while maintaining comparable or superior effectiveness, generating a concise set of semantic types that accurately describe annotated columns.
    \item Model Robustness: 
    performance remains consistent across different commercial and open-source LLMs, demonstrating that our framework is not dependent on a specific model.  
    \item Multi-Type Annotation: StraTyper significantly outperforms LLM-only baselines in identifying and annotating multi-typed columns, achieving $\sim$36\% higher recall while maintaining precision.
    \item Downstream Task Effectiveness: Semantic types discovered by StraTyper yield $\sim$21\% and $\sim$10\% higher accuracy for join discovery and schema matching, respectively, compared to the best-performing LLM-only baseline, validating the quality and utility of discovered types.
\end{itemize}

\subsection{Experiment Setup}
\label{ssec: expsetup}

\begin{table}[t!]
\centering
\footnotesize
\resizebox{\columnwidth}{!}{
\begin{tabular}{l|c||c||c||c||c}
\toprule\bottomrule
\textbf{Dataset} & \textbf{\#Tables} & \begin{tabular}[c]{@{}c@{}}\textbf{Min - Max}\\ \textbf{\#Cols}\end{tabular}  & \begin{tabular}[c]{@{}c@{}}\textbf{Min - Max}\\ \textbf{\#Rows}\end{tabular}&\begin{tabular}[c]{@{}c@{}}\textbf{\#Numerical}\\ \textbf{Cols}\end{tabular}  & \begin{tabular}[c]{@{}c@{}}\textbf{\#Textual}\\ \textbf{Cols}\end{tabular}\\ \toprule\bottomrule
\textbf{NYC-E} & 373 &  2 - 19 & 3 - 1271726 & 1965 & 1444 \\ \hline
\textbf{NYC-CG} & 331 & 3 - 20 &  2 - 100000 & 1378 & 1693 \\ \hline
\textbf{HHS-US} & 97 & 3 - 20 &  5 - 985608 & 456 & 505 \\
\toprule\bottomrule
\end{tabular}
}
\caption{Statistics of the evaluation dataset collections used for column type discovery and annotation in our paper. }
\label{tab:datasets}
\end{table}

\begin{table}[t!]
    \centering
    \footnotesize
    \resizebox{\columnwidth}{!}{
    \begin{tabular}{l | l || l}
    \toprule \bottomrule
    \textbf{Component} & \textbf{Parameter} & \textbf{Setting} \\
    \toprule \bottomrule
    \multirow{4}{*}{Discovery} & Model & Gemini 3 Flash \\ \cline{2-3}
                               & Temperature & 0.3 \\ \cline{2-3}
                               & Value Samples & 10 \\ \cline{2-3}
                               & Context Columns & 5 \\ 
    \hline
    \multirow{2}{*}{Clustering} & Histogram Bins  & 20 \\ \cline{2-3}
                                & Embedding Model & BAAI/bge-base-en-v1.5\\
    \hline
    \multirow{2}{*}{Annotation} & Model & Open AI GPT-OSS \\ \cline{2-3}
                                & Temperature & 0.0 \\
    \hline
    \multirow{2}{*}{Evaluation} & Judge Model & OpenAI GPT-5.4-Mini \\ \cline{2-3}
                                & Temperature & 0.0 \\
    \hline
    \multirow{2}{*}{General}    & Table Context & 3 rows (10 for LLM baselines) \\ \cline{2-3}
                                & Minimum Cluster Size & 2 columns \\
    \toprule \bottomrule
    \end{tabular}}
    \caption{Standard hyperparameters and model settings for \method's variants.}
    \label{tab:hyperparameters}
\end{table}

\para{Dataset Collections} We evaluate \method{} on three diverse dataset collections with different characteristics (Table~\ref{tab:datasets}). Two collections are from the New York City OpenData repository~\cite{nycopendata}: the Education (NYC-E) and City Government (NYC-CG) categories, which represent domain-specific, real-world data. We also include HHS-US, a collection of 97 CSV datasets published by the U.S. Department of Health and Human Services on Data.gov\footnote{\url{https://catalog.data.gov/organization/hhs-gov?res\_format=CSV\&\_tags\_limit=0\&tags=united-states}}, filtered by the tag ``united-states''. These three collections exhibit complementary characteristics, spanning a wide range of table sizes and column types and including both numerical and textual values. NYC-E is predominantly numerical; NYC-CG contains more textual columns. At the same time, HHS-US shows a nearly balanced distribution between numerical and textual attributes and introduces a federal government domain distinct from the municipal focus of the NYC collections. The collections also differ in their degree of semantic redundancy, that is, the extent to which columns across datasets share common semantic types. NYC-E and HHS-US exhibit high semantic redundancy, with many columns reusing the same underlying concepts across datasets. In contrast, NYC-CG shows lower redundancy, with a more diverse set of semantic types and fewer columns per type. This diversity enables us to assess \method{}'s robustness across different data distributions, semantic type categories, and levels of semantic redundancy. We deliberately do not evaluate on standard CTA benchmarks such as SOTAB \cite{korini2022sotab}. As Feuer et al.~\cite{feuer2024archetype} also observed, such benchmarks evaluate classification over fixed label sets drawn from well-known ontologies such as Schema.org or DBPedia \cite{auer2007dbpedia}, which are considerably coarser than the domain-specific, fine-grained types that \method{} is designed to discover. 
Additionally, tables in these benchmarks are artificially stripped of their original column names, since the ground truth labels were originally derived by matching those headers to ontology types~\cite{hulsebos2019sherlock, korini2022sotab}. Keeping headers would make the task trivially solvable by exact-match lookup. This makes these benchmarks unsuitable for our setting, as \method{} leverages column headers as a primary signal for clustering and type discovery. 

\para{\method{} Variants} To evaluate the impact of similarity thresholds and the effectiveness of Iterative Cascading (IC), we instantiate \method{} using distinct configurations
that share the same architecture but differ in how they handle the discovery phase:

\noindent-- \emph{\method$_{Threshold}$}  applies a fixed similarity threshold $\tau_{name} = \tau_{value} = \tau$ for both the name-based and value-based clustering steps. We examine thresholds $\tau \in \{0.75, 0.9, 0.99\}$, ranging from relaxed (broader coverage, potentially lower precision) to strict (higher precision, narrower coverage).   

\noindent-- \emph{\method$_{IC}$} utilizes the iterative cascading discovery strategy from \Cref{sec:iterative-cascading}, operating over the threshold sequence $\{0.99, 0.9, 0.75\}$. This variant eliminates the trade-off inherent in selecting a single fixed threshold, effectively balancing high precision with broad coverage. We adopt this as the default variant of \method.

Unless otherwise noted, all experiments adhere to the standard parameter settings detailed in \autoref{tab:hyperparameters}. 

\para{Model Selection} We select specific Large Language Models (LLMs) and temperature settings tailored to the distinct requirements of each phase (creativity for discovery vs. determinism for classification and evaluation). 
%
For discovery, we use Gemini 3 Flash for its high inference throughput,
encapsulated knowledge, and performance comparable to larger models (e.g., Gemini Pro or Anthropic's premier models). 
For closed-set annotation, we employ the open-source OpenAI GPT-OSS, as with our method's structured prompting 
strategy, open-source models perform on par with commercial ones. Other open-source models like Microsoft's Phi-4 can be substituted without performance degradation. 
%
To ensure a fair, automated evaluation, we use OpenAI GPT-5.4-Mini as a judge, as this model family has been shown to perform best in LLM-as-Judge tasks~\cite{zheng2023judging, liu2023g}.

\para{Baselines} Existing CTA methods \cite{hulsebos2019sherlock, zhang2020sato, deng2021turl, miao2023watchog, suhara2022annotating,korini2023column, kayali2024chorus,li2024table, feuer2024archetype, wei2024racoon, ding2025retrieve, xiao2025cents} address a different problem than \method{}: they assume a predefined type vocabulary, typically drawn from generic ontologies (e.g., Schema.org or DBpedia)~\cite{auer2007dbpedia}, and treat annotation as classifying each column into one of the types, whereas \method{} discovers the vocabulary itself, tailored to the input collection, and handles multi-typed columns. While a direct comparison is therefore not ideal, we evaluate two representative CTA methods, Sato~\cite{zhang2020sato} and Doduo~\cite{suhara2022annotating}, both pretrained on VizNet~\cite{hu2019viznet} over a fixed vocabulary of 78 semantic types, on all three collections with the LLM-as-Judge protocol described below. 
\begin{figure*}[t!]
    \centering
    \includegraphics[scale=0.28]{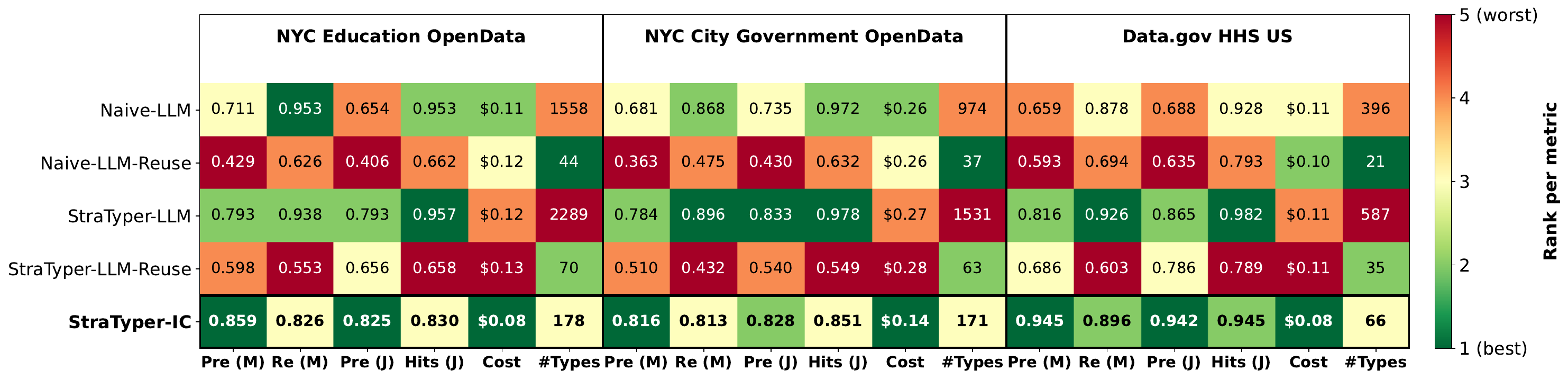}
    \caption{Effectiveness vs. Cost vs. Type Redundancy trade-off across three datasets. Pre refers to Precision, Re to Recall, (M) and (J) refer to Manual and LLM-Judge evaluations respectively, while Cost is aggregated per thousand columns annotated.}
    \label{fig:heatmap}
\end{figure*}
Our primary baselines are LLM-driven approaches that can operate without a predefined vocabulary. We design four variants that rely exclusively on an LLM to isolate the contributions of \method's 
prompt instructions and type-reuse mechanism.

\noindent-- \emph{\methodllm{}} applies \method{}'s prompt instructions (\Cref{sec:discovery}) to every dataset in the collection to annotate their columns. This allows us to assess the efficacy of our structured prompting in isolation from our clustering and cascading mechanisms. Given that this baseline combines our elaborate type-discovery instructions with a powerful commercial LLM, it essentially functions as an oracle, i.e., an upper bound on performance, representing the ideal quality achievable when resource constraints (e.g., cost) are removed.

\noindent-- \emph{\methodllm{Reuse}} extends \methodllm{} by maintaining a dynamic list of 
discovered types and providing instructions on when and how to reuse them. 
We use this baseline to assess whether the LLM can consistently select valid types from an evolving list without our inverted-index structure.

\noindent-- \emph{\naive{}} represents a straightforward approach to the problem studied in this paper. It 
prompts the LLM to identify all distinct possible semantic types for all columns of each given dataset without further instructions and guidelines. We employ this baseline to assess the effectiveness of our structured prompt design. 

\noindent-- \emph{\naive{Reuse}} combines the  prompt of \naive{} with the type reuse mechanism of \methodllm{Reuse}. It provides the LLM with a list of previously discovered types but offers no explicit guidance on when or how to select them.

To ensure a fair and consistent comparison,  the LLM-driven baselines utilize Gemini 3 Flash and are given the same input context: the column headers and a random sample of 10 data rows, filtered to minimize \texttt{NaN} values and maximize information density.

\para{Automated Evaluation with LLM-as-a-Judge} Manual verification of semantic type annotations across hundreds of columns is prohibitively expensive. Following recent work demonstrating the reliability of LLM-based evaluation for reasoning~\cite{zheng2023judging}, natural language generation~\cite{liu2023g}, and dataset description~\cite{zhang2025autoddg} tasks, we use an LLM as a judge. 

For each column, the judge receives table context (column names and sampled rows), additional values from the column, and the predicted annotations of all methods. We pool these predictions into a single set of unique candidate types, so that identical types receive the same verdict regardless of which method produced them. The judge marks an annotation as correct if the type accurately describes a subset of the column's values, which accommodates multi-typed columns. Because we target rich types that refer to real-world concepts, the judge rejects generic types (e.g., ``number'', ``string'').

{We use two metrics. \emph{Hit Rate} is the fraction of columns for which a method provides at least one correct annotation, where $\mathbb{I}(\cdot)$ is the indicator function:
\begin{equation}
Hits = \frac{1}{|C|} \sum_{c \in C} \mathbb{I}\left(|T_{ann}(c) \cap T_{valid}(c)| \ge 1\right).
\end{equation}
\emph{Precision} is the fraction of annotated types that are correct:
\begin{equation}
P_{judge} = \frac{\sum_{c \in C} |T_{ann}(c) \cap T_{valid}(c)|}{\sum_{c \in C} |T_{ann}(c)|},
\end{equation}
where $C$ is the set of annotated columns, and $T_{ann}(c)$ and $T_{valid}(c)$ are the types annotated for column $c$ and judged correct, respectively.


\para{Manual Evaluation on Stratified Samples} While automated evaluation scales effectively across all annotated columns, it cannot identify valid types that a method failed to discover (false negatives). 
Therefore, we conduct a manual evaluation on a stratified subset of columns. We use \method$_{IC}$ to guide sampling by selecting one representative column for each distinct type discovered, ensuring full coverage of the captured semantic space. 

For each sampled column and each method, we construct a synthetic ground truth ($GT_{syn}$) of valid semantic types through a two-step process: 1) \emph{Validate predicted types:} A human evaluator inspects each method's annotations and adds  correct semantic types to the ground truth $GT_{syn}$, and
2) \emph{Identify missed types:} The evaluator manually reviews the column's values and table context to identify valid semantic concepts that all methods failed to discover. These missed concepts are added to $GT_{syn}$ as abstract placeholders (e.g., \texttt{Other$_1$}, \texttt{Other$_2$}). To ensure annotation quality, each examined column was studied thoroughly: the evaluator inspected the full list of unique values, any available column and dataset descriptions, the surrounding table context, and additional metadata provided by the source repository. This thorough inspection ensures that the ground truth reflects a well-informed understanding of each column's semantics rather than a superficial judgment based on a few sampled values. Including the \texttt{Other$_k$} placeholders ensures that methods are penalized for false negatives, enabling calculation of Recall and F1 scores. 

A key challenge is that methods may propose correct types at different granularities: one method might produce ``NYC DBN'' while another produces ``School Identifier'' for the same column. To properly capture the difference in granularity, we distinguish between correct and desired target annotations. Methods that miss the finer-grained type are penalized only on recall (by adding it to the $GT_{syn}$), not precision.


This evaluation strategy allows computing true-positive (TP), false-positive (FP), and false-negative (FN) counts over the sampled columns $S$ as $TP = \sum_{c \in S} |T_{ann}(c) \cap GT_{syn}(c)|$, $FP = \sum_{c \in S} |T_{ann}(c) \setminus GT_{syn}(c)|$, and $FN = \sum_{c \in S} |GT_{syn}(c) \setminus T_{ann}(c)|$, where $T_{ann}(c)$ is the set of semantic type annotations for a column $c \in S$. We therefore compute Precision, Recall, and F1 over all sampled columns.

\subsection{Effectiveness vs. Cost vs. Type Redundancy}
\label{sec:evaluation}

\Cref{fig:heatmap} summarizes annotation quality, LLM usage cost, and type redundancy across all methods and dataset collections. Each cell reports the raw metric value, shaded by the method's rank on that metric (green for best, red for worst).
The four leftmost columns capture quality (precision and recall computed based on manual evaluation, plus LLM-judge precision and hit rate), and the two rightmost columns capture cost-efficiency and conciseness 
(cost per thousand columns annotated and number of unique types discovered).  

\para{Comparison of Methods}  \method$_{IC}$ balances all three dimensions in a way no LLM-only baseline matches: it is consistently the most cost-efficient method and produces a compact yet meaningful type vocabulary, avoiding both the excess of non-reuse baselines (from 587 to 2,289 types) and the over-compression of baselines that employ type reuse (as few as 21 types). This efficiency does not come at the expense of quality, since \method$_{IC}$ delivers the most precise annotations across all three collections under both manual and judge-based evaluation, as it propagates discovered types within seed clusters and scopes closed-set annotation to small, high-confidence candidate sets. Recall and hit rate remain among the highest, comparable to those of LLM-only baselines that process each column with a powerful proprietary model. 

\begin{table}[t!]
    \centering
    \renewcommand{\arraystretch}{1.0}
    \footnotesize
    \setlength{\tabcolsep}{4pt}
    \resizebox{\columnwidth}{!}{
    \begin{tabular}{l | c | c || c | c || c | c}
    \hline\hline
    \multirow{2}{*}{\textbf{Method}} & \multicolumn{2}{c||}{\textbf{NYC-E}} & \multicolumn{2}{c||}{\textbf{NYC-CG}} & \multicolumn{2}{c}{\textbf{HHS-US}} \\ \cline{2-7}
     & \textbf{Pre} & \textbf{Hits} & \textbf{Pre} & \textbf{Hits} & \textbf{Pre} & \textbf{Hits} \\
    \hline\hline
    Sato \cite{zhang2020sato} & 0.244 & 0.244 & 0.349 & 0.349 & 0.248 & 0.248 \\ \hline
    Doduo \cite{suhara2022annotating} & 0.248 & 0.248 & 0.344 & 0.344 & 0.157 & 0.157 \\ \hline
    \textbf{\method$_{IC}$} & \textbf{0.825} & \textbf{0.830} & \textbf{0.828} & \textbf{0.851} & \textbf{0.942} & \textbf{0.945} \\
    \hline\hline
    \end{tabular}}
    \caption{Judge-based Precision (Pre) and Hit Rate (Hits) of CTA methods compared to \method$_{IC}$.}
    \label{tab:cta_comparison}
\end{table}

The oracle \methodllm{}, which applies our structured prompt without clustering or cascading, consistently ranks first or second on every quality metric under both evaluations, yet at a steep price: its vocabulary grows to 2,289 types on NYC-E and 1,531 on NYC-CG, roughly an order of magnitude more than that of \method$_{IC}$, because the model independently generates synonymous types for similar columns across datasets, and its cost is up to twice that of \method$_{IC}$. \naive{} exhibits a different failure mode: while its hit rate and recall are high, its precision is substantially lower, since without explicit guidance the model frequently proposes generic types (e.g., ``numerical\_value'') that the judge may accept but that fail to capture domain-specific semantics, which makes it unreliable for downstream tasks that require consistent and precise annotations. Finally, the reuse baselines \naive{Reuse} and \methodllm{Reuse} produce the smallest vocabularies, yet they collapse in quality on all collections; manual inspection reveals that once a generic type enters the reuse list, the model prioritizes it for subsequent columns, consistent with previous findings that LLMs struggle to select correct types as candidate lists grow~\cite{feuer2024archetype}. This global, ever-growing candidate list differs fundamentally from our inverted-index-guided reuse, which surfaces only the small subset of candidates relevant to each cluster and column.

 \para{LLM-judge vs. Manual Evaluation} A direct comparison of judge-based and manual metrics across the heatmap reveals strong agreement between the two strategies. Precision values are closely aligned for every method: differences rarely exceed a few percentage points, and the relative ranking of methods is preserved across both evaluations. The same holds for coverage-oriented metrics, hit rate and recall, which, though measured against different ground truths, produce comparable values and near-identical method rankings. This consistency indicates that the LLM-based judge can be trusted as a scalable proxy for human evaluation, reproducing the conclusions that manual inspection would yield. Manual evaluation confirms the validity of our large-scale automated evaluation, with the few observed differences primarily attributable to the judge's limited context relative to a human evaluator with full data access. This reinforces the utility of LLM-based judges for scaling evaluation while highlighting the need to pair them with manual verification to capture nuances.

\subsection{Comparison Against CTA Methods} \label{ssec:cta}
 
We evaluate Sato \cite{zhang2020sato} and Doduo \cite{suhara2022annotating}, pretrained on VizNet~\cite{hu2019viznet} with their fixed vocabularies of 78 semantic types, on all three dataset collections, and assess their annotations with the LLM-as-Judge protocol of Section~\ref{ssec: expsetup}, reporting 
Precision and Hit Rate. Table~\ref{tab:cta_comparison} presents the results.
Note that, since both methods assign exactly one type per column, their precision and hit rate coincide. As expected, both perform poorly across all collections, as their pretrained vocabularies cannot express the domain-specific semantic types these collections require, leaving the majority of columns incorrectly annotated. Moreover, the columns they do annotate correctly typically receive generic types covered by the VizNet vocabulary, such as names, addresses, or descriptions, rather than the fine-grained, collection-specific concepts that downstream tasks need. Their results are strongest (0.349) on NYC-CG, whose textual columns include more generic concepts, and weakest on HHS-US (0.157), which is dominated by specialized federal health semantics. \method$_{IC}$ surpasses both methods by wide margins on all collections (0.825--0.945), underscoring the value of discovering semantic types tailored to a collection before annotating with them.

\subsection{Semantic Types for Column Join/Matches}
\label{sec:downstream}

\newcommand{\freyjacol}{JD}
\newcommand{\nycecol}{SM}
\begin{table}[t]
\centering
\renewcommand{\arraystretch}{1.0}
\setlength{\tabcolsep}{3pt}
\newlength{\datacol}
\setlength{\datacol}{1.7cm}
\resizebox{\columnwidth}{!}{%
\begin{tabular}{l|>{\centering\arraybackslash}p{\datacol}|>{\centering\arraybackslash}p{\datacol}||>{\centering\arraybackslash}p{\datacol}|>{\centering\arraybackslash}p{\datacol}}
\toprule \bottomrule
\multirow{2}{*}{\textbf{Method}} & \multicolumn{2}{c||}{\textbf{Avg. Emb. Similarity}} & \multicolumn{2}{c}{\textbf{\% Pairs with Same Types}} \\ \cline{2-5}
 & \textbf{\freyjacol} & \textbf{\nycecol} & \textbf{\freyjacol} & \textbf{\nycecol} \\ \toprule \bottomrule
\naivebold{} & 0.788 & 0.843 & 7.36 & 9.14 \\ \hline
\naivebold{Reuse} & 0.930 & 0.862 & 60.11 & 41.08 \\ \hline \hline
\methodllmbold{} & 0.773 & 0.823 & 9.56 & 7.79 \\ \hline
\methodllmbold{Reuse} & 0.822 & 0.894 & 48.79 & 74.50\\ \hline \hline
\textbf{\method{}$_{IC}$} & \textbf{0.936} & \textbf{0.949} & \textbf{81.32} & \textbf{84.28} \\ 
\toprule \bottomrule
\end{tabular}%
}
\caption{Semantic type similarity for join discovery (JD) and schema matching (SM) using Freyja and NYC-E column pairs, respectively.}
\label{tab:match-types}
\end{table}

We study the effectiveness of \method{} 
for two downstream tasks: \emph{join discovery} \cite{dong2023deepjoin, maynou2024freyja, guo2025snoopy, koutras2025omnimatch, liu2026hyperjoin} and \emph{schema matching} \cite{rahm2001survey, koutras2021valentine, liu2025magneto}.

\para{Join/Match Ground Truth Pairs}  To evaluate \method's annotations, we need ground truth join/match column pairs. We utilize two distinct ground truth datasets. For join discovery, we use the Freyja benchmark \cite{maynou2024freyja}, which contains 910 equi-join and fuzzy-join pairs. For schema matching, we curated new ground truth for NYC-E
by manually identifying semantically corresponding column pairs. We used \methodllm{} semantic types (the highest quality, per \Cref{sec:evaluation}) to facilitate annotation: we computed clusters of columns with semantically similar types, sampled column pairs from each cluster, and manually verified them as valid matches. Matches were accepted if they represented valid equi-join, fuzzy-join, or union (same semantics, no overlaps) cases. Based on this process, we constructed a ground truth of 1,412 matches.

\para{Evaluation Process} For each pair of join/match columns, we compute the average embedding similarity of their annotations, which indicates how semantically similar the semantic types of joinable/matching columns are. In addition, we compute the percentage of column pairs that have exactly the same type annotation sets, which provides stronger evidence of whether a semantic type annotation method is suitable for capturing joinable/matching pairs. \Cref{tab:match-types} summarizes the results for \method{} and LLM-only baselines for join discovery (Freyja) and schema matching (NYC-E).

\para{Join Discovery} \method{} provides the most useful semantic types for join discovery: joinable pairs exhibit the highest embedding similarity, and over 81\% receive the same type annotations, substantially outperforming all baselines. Without type reuse, methods assign identical types to fewer than 10\% of joinable pairs, highlighting the severity of type redundancy when LLMs generate types without a curated set of candidates. These results suggest that \method{} can be seamlessly integrated into any join discovery method as a filtering step, where a simple check for annotation equality between column pairs might suffice. Baselines that reuse discovered types reduce redundancy by forcing the LLM to select from a closed set. Yet, this naive reuse strategy is far less effective than our iterative cascading discovery process, which achieves substantially better annotation consistency.

\para{Schema Matching} \method's semantic types are also the most reliable for determining schema matches, achieving the highest embedding similarity and percentage of pairs with identical annotations. The latter result is particularly noteworthy: despite the schema-matching ground truth being more diverse than join discovery, including fuzzy-join and union scenarios where column pairs share semantics but may have partially overlapping or non-overlapping value sets, \method$_{IC}$ outperforms all LLM-only baselines. This confirms that iterative cascading
enables it to capture the broader semantic relationships present in complex schema-matching tasks. As in join discovery, type reuse is critical: baselines without reuse assign identical types to fewer than 10\% of matching pairs, whereas \method$_{IC}$ reaches 84\%.

\begin{table}[t!]
    \centering
\begin{footnotesize}
    \renewcommand{\arraystretch}{1.0}
    \setlength{\tabcolsep}{4pt}
    \resizebox{\columnwidth}{!}{
    \begin{tabular}{l | l | l | c | c | c | c}
    \hline\hline
    \multirow{2}{*}{\textbf{Dataset}} & \multirow{2}{*}{\textbf{Stage}} & \multirow{2}{*}{\textbf{Data Type}} & \multicolumn{3}{c|}{\textbf{\method$_{IC}$}} & \multirow{2}{*}{\textbf{\# Cols}} \\ \cline{4-6}
     & & & \textbf{Precision} & \textbf{Recall} & \textbf{F1} & \\
    \hline\hline
    
    \multirow{4}{*}{\textbf{NYC-E}}
      & \multirow{2}{*}{Discovery}  & Numerical & 0.858 & 0.858 & 0.858 & 106 \\ \cline{3-7}
      &                             & Textual   & 0.897 & 0.824 & 0.859 & 64 \\ \cline{2-7}
      & \multirow{2}{*}{Annotation} & Numerical & 0.826 & 0.826 & 0.826 & 69 \\ \cline{3-7}
      &                             & Textual   & 0.843 & 0.768 & 0.804 & 44 \\
    \hline
    
    \multirow{4}{*}{\textbf{NYC-CG}}
      & \multirow{2}{*}{Discovery}  & Numerical & 0.933 & 0.933 & 0.933 & 60 \\ \cline{3-7}
      &                             & Textual   & 0.882 & 0.866 & 0.874 & 100 \\ \cline{2-7}
      & \multirow{2}{*}{Annotation} & Numerical & 0.889 & 0.889 & 0.889 & 45 \\ \cline{3-7}
      &                             & Textual   & 0.640 & 0.646 & 0.643 & 87 \\ 
    \hline
    
    \multirow{4}{*}{\textbf{HHS-US}}
      & \multirow{2}{*}{Discovery}  & Numerical & 0.941 & 0.941 & 0.941 & 34 \\ \cline{3-7}
      &                             & Textual   & 0.941 & 0.821 & 0.877 & 30 \\ \cline{2-7}
      & \multirow{2}{*}{Annotation} & Numerical & 0.933 & 0.933 & 0.933 & 15 \\ \cline{3-7}
      &                             & Textual   & 0.962 & 0.926 & 0.943 & 17 \\
    \hline\hline
    \end{tabular}}
    \caption{Detailed manual evaluation of \method$_{IC}$ across discovery and annotation stages, broken down by numerical and textual columns.}
    \label{tab:detailed_analysis}
\end{footnotesize}
\end{table}

\subsection{Annotation Quality: Detailed Analysis}

Table~\ref{tab:detailed_analysis} reports  manually evaluated precision, recall, and F1 of \method$_{IC}$ across collections, broken down by stage and column data type, revealing how pipeline stage and column structure affect quality.
%
Discovery-stage annotations are higher quality than those produced during the annotation stage. This is expected: discovery operates over seed clusters via a proprietary LLM, benefiting from coherent groupings and strong model capability (Section \ref{sec:seed-cluster-propagation}), while the annotation stage handles residual columns via closed-set selection with an open-source LLM -- a harder task even for closed-source models.
Manual inspection shows a characteristic failure mode: when no candidate accurately describes a column, the LLM occasionally picks a semantically adjacent type rather than abstaining, despite our instructions to reject non-matching candidates.

Numerical columns are annotated more reliably than textual ones for two reasons: numerical columns carry a single semantic type, while textual columns may span multiple types, and 
numerical semantics are more often anchored by column name and table context, whereas textual ambiguity requires finer-grained, value-level reasoning.  
The discovery-to-annotation quality drop is accordingly most pronounced for textual columns. 

Finally, semantic redundancy across a collection’s constituent datasets shapes outcomes. NYC-E and HHS-US exhibit high redundancy, letting seed clusters capture most columns and propagate their types. NYC-CG instead contains many long-tail columns that do not form seed clusters and fall back to closed-set annotation with an open-source model — a choice we use to keep costs low, but one that is not fixed: a commercial LLM can be substituted here without changing the framework when cost is not a constraint.


\subsection{Ablation Studies}
\label{sec:ablation}
\para{Model Ablation} We evaluate \method's sensitivity to the underlying LLM on NYC-E using Microsoft's Phi-4 (14B, open-source), OpenAI's GPT-OSS (120B, open-source), and commercial Gemini 3 Flash. \Cref{fig:ablation_combined} shows annotation accuracy and discovered type counts for each model across the discovery and annotation phases.

\begin{figure}[t!]
    \centering
    \includegraphics[width=.95\columnwidth]{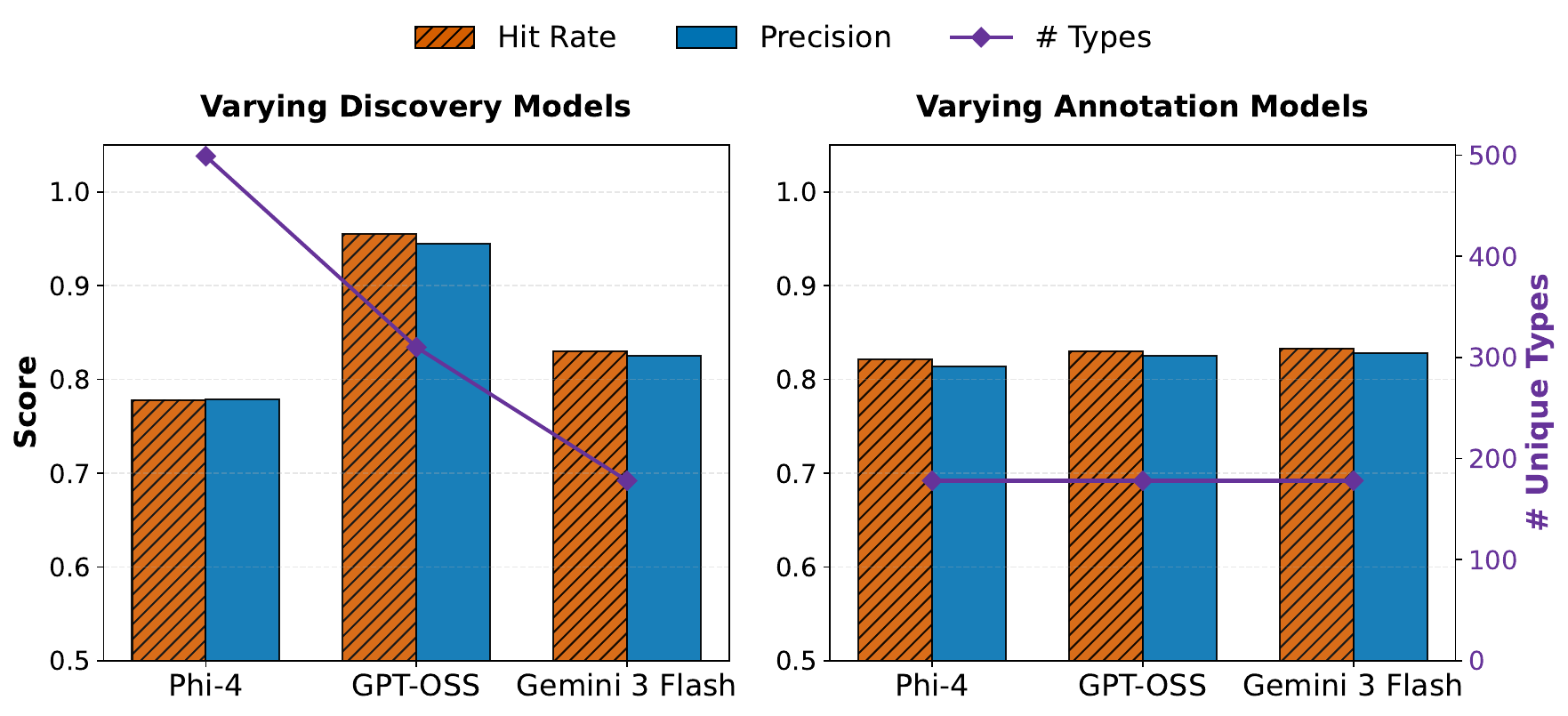}
    \caption{Impact of varying the underlying LLM for the discovery and annotation phases. Bars denote effectiveness metrics, while the line tracks number of unique types discovered.}
    \label{fig:ablation_combined}
\end{figure}

Model choice significantly affects discovery quality, though not always in ways automated metrics capture. Phi-4 achieves a hit rate comparable to larger models but with lower precision; surprisingly, GPT-OSS scores highest on both hit rate and precision. However, manual inspection reveals why: open-source models frequently create synonymous types for identical concepts or trivially mirror column names, and the LLM judge may still score these as correct, inflating effectiveness metrics. Gemini 3 Flash instead generates more diverse, semantically meaningful types, motivating our default choice for discovery.

During annotation, all three models perform nearly identically (within 1\%), showing that \method's concise candidate sets let open-source models match proprietary ones without sacrificing accuracy. The number of unique types also remains stable across models, confirming that closed-set annotation prevents hallucinating types outside the candidate set -- a common failure mode when LLMs generate rather than select types.

\begin{table}[t!]
    \centering
    \renewcommand{\arraystretch}{1.0}
    \footnotesize
    \resizebox{\columnwidth}{!}{
    \begin{tabular}{l|c|c|c|c}
    \toprule \bottomrule
    \textbf{Method} & \textbf{Hit Rate} & \textbf{Precision} & \textbf{\# Unique Types} & \textbf{Cov.} \\ 
    \hline\hline
    \textbf{\method}$_{75}$  & 0.889 & 0.886 &  139 & 82.37\%\\ \hline
    \textbf{\method}$_{90}$  & \textbf{0.896} & \textbf{0.892} & 157 & 75.36\%\\ \hline
    \textbf{\method}$_{99}$  & 0.870 & 0.869 & 141 & 66.24\%\\ \hline
    \textbf{\method}$_{IC}$  & 0.830 & 0.825 & 178 & 83.90\%\\ 
    \toprule \bottomrule
    \end{tabular}}
    \caption{Comparison of \method's variants.}
    \label{tab:stratyper-variants}
\end{table}

\para{Iterative Cascading vs.\ Single-Threshold Discovery} Table~\ref{tab:stratyper-variants} compares \method$_{IC}$ against single-threshold variants on NYC-E. Hit rate and precision vary little across configurations; stricter thresholds give modest precision gains by forming tighter clusters, but at substantial coverage cost, since fewer, more specific types apply to fewer columns as $\tau$ increases. \method$_{75}$ achieves the highest coverage among the single-threshold variants, but manual inspection shows this comes from over-generalization: looser clustering groups semantically distinct columns together, forcing overly generic types (e.g., ``Location'' instead of ``NYC Borough''). \method$_{IC}$ avoids this by discovering types at multiple granularities, maintaining robust coverage with a vocabulary that is both distinct and meaningful.

\begin{table}[t!]
    \centering
    \footnotesize
    \renewcommand{\arraystretch}{1.0}
    \begin{tabular}{l|c|c|c}
    \toprule \bottomrule
    \textbf{Method} & \textbf{Precision} & \textbf{Recall} & \textbf{F1} \\ 
    \hline\hline
    \methodllmbold{} & 0.916 & 0.510 & 0.655 \\ \hline \hline
    \textbf{\method}$_{IC}$  & \textbf{0.983} & \textbf{0.694} &  \textbf{0.814}\\ 
    \toprule \bottomrule
    \end{tabular}
    \caption{Comparison on multi-typed columns. }
    \label{tab:multitype}
\end{table}

\para{Annotating Multi-Type Columns} We manually evaluated the complete set of 30 multi-typed columns identified in NYC-E (\Cref{tab:multitype}), comparing \method$_{IC}$ to the oracle-like \methodllm{}. Both achieve high precision, but \method$_{IC}$'s recall is substantially higher, capturing more valid types when a column spans several domains, a gap attributable to its dedicated value sampling strategy, since both methods share the same discovery instructions.

\para{Value Sampling} We compare length-stratified sampling against two clustering-based alternatives (TF-IDF and SBERT value embeddings, both clustered via k-means into as many clusters as the sampling budget) over NYC-E's multi-typed columns, measuring the fraction of semantic types each strategy represents with at least one sample. Length-stratified sampling reaches near-perfect coverage (98.8\%), versus 84.0\% and 81.5\% for TF-IDF and SBERT, since both baselines let multiple clusters cover the same dominant type, starving rarer ones of budget. String length instead tracks type boundaries in our collections, guaranteeing a representative per structural group at negligible cost; sampling remains modular, and combining structural and semantic signals is left for future work.

\para{Seed Cluster Precision} \method{} assumes high-quality seed clusters. We validate this by sampling five clusters per threshold $\tau \in \{0.99, 0.9, 0.75\}$ and collection, stratified across small (2--5 columns), medium (6--10), and large ($>$10) clusters, manually labeling every column's full type set, and computing the fraction of within-cluster pairs sharing at least one type and the fraction with identical type sets. As Table~\ref{tab:seed_precision} shows, clusters at $\tau=0.99$ are perfectly pure everywhere; at $\tau=0.9$, every pair shares at least one type; and deviations appear only at $\tau=0.75$, where the large majority of pairs still agree. The remaining impurities stem from multi-typed columns with partially overlapping types or from semantically adjacent columns (e.g., first and last names) that share no type. These results confirm the intersection's role as a high-precision filter and motivate Iterative Cascading's strategy of forming pure clusters first and propagating their types at looser thresholds.

\begin{table}[t]
    \centering
    \renewcommand{\arraystretch}{1.0}
    \setlength{\tabcolsep}{4pt}
    \resizebox{\columnwidth}{!}{
    \begin{tabular}{l | c | c | c | c | c | c}
    \hline\hline
    \textbf{Dataset} & \textbf{$\tau$} & \textbf{\#Clusters} & \textbf{\#Cols} & \textbf{\#Pairs} & \textbf{$\geq$1 Shared (\%)} & \textbf{Identical (\%)} \\
    \hline\hline
    \multirow{3}{*}{\textbf{NYC-E}}
      & 0.99 & 15 & 199 & 3,927 & 100.0 & 100.0 \\ \cline{2-7}
      & 0.90 & 15 & 249 & 9,290 & 100.0 & 100.0 \\ \cline{2-7}
      & 0.75 & 15 & 134 & 907 & 100.0 & 99.0 \\
    \hline
    \multirow{3}{*}{\textbf{NYC-CG}}
      & 0.99 & 10 & 60 & 281 & 100.0 & 100.0 \\ \cline{2-7}
      & 0.90 & 15 & 155 & 1,303 & 100.0 & 100.0 \\ \cline{2-7}
      & 0.75 & 15 & 175 & 2,006 & 99.1 & 94.0 \\
    \hline
    \multirow{3}{*}{\textbf{HHS-US}}
      & 0.99 & 13 & 104 & 735 & 100.0 & 100.0 \\ \cline{2-7}
      & 0.90 & 15 & 137 & 889 & 100.0 & 99.3 \\ \cline{2-7}
      & 0.75 & 15 & 129 & 741 & 98.8 & 98.0 \\
    \hline\hline
    \end{tabular}}
    \caption{Seed cluster precision under manual ground truth.}
    \label{tab:seed_precision}
\end{table}

\section{Related Work}

\para{Column Type Annotation (CTA)} 
CTA is typically modeled as classification over a predefined set of semantic types. Seminal works such as Sherlock \cite{hulsebos2019sherlock} and Sato \cite{zhang2020sato} train deep learning models on large labeled corpora, while others \cite{chen2019colnet, chen2019learning, takeoka2019meimei, su2023sand, langenecker2024pythagoras, khorram2025mixedsand} pair such models with Knowledge Bases for entity-level metadata. These methods require large labeled datasets and closed, often KB-limited, vocabularies, and assign a single type per column. AdaTyper \cite{hulsebos2023adatyper} adapts a pretrained predictor to new types via user corrections, but still assumes an initial closed vocabulary, human feedback, and single-type columns. More recent work uses LLMs for zero-shot CTA given a predefined type set \cite{korini2023column,kayali2024chorus,feuer2024archetype}. Unlike these approaches, \method{} targets open-world, multi-type annotation without predefined labels or training data, combining commercial LLMs for zero-shot discovery with open-source models for closed-set annotation.

\para{Concept/Domain Discovery} C$^4$~\cite{li2017discovering} discovers concept hierarchies from spreadsheets by constructing a graph where edges represent the Jaccard similarity between terms, filtering weak links to identify connected components as concepts; but it relies on user-defined thresholds and assumes high column homogeneity. D$^4$~\cite{ota2020data} improves polysemy handling (e.g., ``black'' as color vs. surname) through aggregated co-occurrence signatures, but remains syntactic and requires tuning multiple hyperparameters.
RaF-STD \cite{piai_fine-grained_2023} discovers fine-grained types for heterogeneous web sources by iteratively combining Bayesian overlap analysis, NLP-inspired tagging, and clustering, but assumes entity-centric data with attribute name-value pairs and relies on cross-source value overlap.  \method{} instead uses semantic embeddings to capture  conceptual relationships that co-occurrence statistics miss, and LLMs to generate interpretable type definitions rather than unlabeled term clusters, with Iterative Cascading removing the dependence on manual single-threshold tuning.

\para{Table Understanding} 
Several table understanding approaches \cite{deng2021turl, suhara2022annotating, wang2023sudowoodo, miao2023watchog, tu2023unicorn, sun2023reca, ding2025retrieve} use encoding techniques to generate representations and fine-tune them in a (semi-)supervised way for column-oriented tasks, such as CTA. Similar models (e.g., \cite{herzig2020tapas, yin2020tabert, wang2021tuta}) have been applied to other tasks, such as text-to-SQL and question answering. 
Despite the benefits these models bring for tabular representations, applying them to CTA still requires fine-tuning on large training corpora and predefined sets of target semantic types.


\para{Related Upstream/Downstream Tasks} Approaches that capture relatedness among columns, such as schema matching \cite{rahm2001survey, do2002coma,koutras2021valentine, liu2025magneto} and join discovery \cite{dong2023deepjoin, maynou2024freyja, guo2025snoopy, koutras2025omnimatch, liu2026hyperjoin}, have a direct connection with CTA methods, since semantic types can be important for refining valid column pairs; indeed, in Section \ref{sec:downstream} we observed that \method's annotations can be effectively integrated into such methods, providing better results than LLM-only baselines. Other data management tasks such as dataset discovery~\cite{fernandez2018aurum, nargesian2018table, bogatu2020dataset, dong2021efficient, fan2023semantics, khatiwada2023santos, hu2023automatic}, automated generation of textual descriptions for datasets \cite{zhang2025autoddg}, and unsupervised error detection that builds on column-level semantic typing \cite{chen2025autotest} could also benefit from the richer, domain-tailored semantic types discovered by \method.

\section{Conclusions and Future Work}

We introduced \method, a cost-effective framework  for automated semantic type discovery and multi-type annotation in dataset collections that addresses fundamental limitations of existing CTA methods. 
Unlike existing CTA methods, which require closed-type vocabularies defined at training or inference time and assign a single type per column, \method{} discovers domain-tailored semantic types without predefined labels and correctly handles columns whose values span multiple semantic types. 
Through strategic column clustering, length-stratified value sampling, dynamic type retrieval, and iterative cascading discovery, \method{}  balances type precision with annotation coverage while minimizing LLM costs.
Our evaluation on three real-world dataset collections demonstrates that \method{} achieves high annotation precision, substantially reduces LLM costs compared to commercial baselines, and produces semantic type vocabularies that are both compact and meaningful. Furthermore, the types discovered by \method{} improve downstream data integration tasks, yielding $\sim$21\% and $\sim$10\% higher accuracy for join discovery and schema matching, respectively, compared to the best LLM-only baseline.

\para{Limitations and Future Directions} While \method{} shows significant improvements for multi-type discovery, the task remains challenging; future work could employ more sophisticated value sampling techniques to better uncover types in such cases. Additionally, our reliance on column names for clustering limits applicability when headers are missing or opaque, suggesting a need for future exploration into header-agnostic profiling techniques and alternative clustering algorithms. Finally, we plan to further refine the synthesis of contextual evidence provided to the models to maximize reasoning precision in highly ambiguous scenarios.

\bibliographystyle{IEEEtran}
\bibliography{references} 

\end{document}